# Duplication models for biological networks

Fan Chung [*], Linyuan Lu[*], T. Gregory Dewey† and David J. Galas†[+]

*Department of Mathematics,
University of California at San Diego
La Jolla, California, 92093

†Keck Graduate Institute of Applied Life Sciences
535 Watson Drive,
Claremont, California 91711

[+] corresponding author, david_galas@kgi.edu, phone: 909 607-7487, Fax: 909 607-8598

## Abstract

Are biological networks different from other large complex networks? Both large biological and non-biological networks exhibit power-law graphs (number of nodes with degree $k$, $N(k) \sim k^{-\beta}$) yet the exponents, $\beta$, fall into different ranges. This may be because duplication of the information in the genome is a dominant evolutionary force in shaping biological networks (like gene regulatory networks and protein-protein interaction networks), and is fundamentally different from the mechanisms thought to dominate the growth of most non-biological networks (such as the internet [1-4]). The preferential choice models non-biological networks like web graphs can only produce power-law graphs with exponents greater than 2 [1-4,8]. We use combinatorial probabilistic methods to examine the evolution of graphs by duplication processes and derive exact analytical relationships between the exponent of the power law and the parameters of the model. Both full duplication of nodes (with all their connections) as well as partial duplication (with only some connections) are analyzed. We demonstrate that partial duplication can produce power-law graphs with exponents less than 2, consistent with current data on biological networks. The power-law exponent for large graphs depends only on the growth process, not on the starting graph.

Networks of interactions are fundamental to all biological systems. The interactions among species in ecosystems, the interactions between cells in an organism, and among molecules in a cell are all parts of complex biological networks. There is considerable current interest in networks within the cell - genetic regulatory networks and protein-protein interaction networks, in particular - about which we can now acquire extensive data using new technological advances. The duplication of the information in the genome - genes and their controlling elements – is a central force in evolution and should be determinative of biological networks.

The process of duplication is quite different from the mechanisms thought to dominate the growth of most non-biological networks (such as the internet, social or citation networks [1-5]), which involve the simple addition of nodes with preferential connection to existing nodes. These latter processes only produce power-laws graphs with exponents greater than 2 [1-4,6,8]. A power-law graph is one in which the number of nodes of degree k (the number of edges impinging on a vertex), $N(k)$, has a distribution that follows a power-law: $N(k) \sim k^{-b}$. We present new mathematical results here on the evolution of graphs by different duplication processes. Using a combinatorial-probabilistic approach to analyze both the full duplication of nodes (with all their connections) as well as partial duplication (with only some of their connections), we find that full duplication retains a strong "memory" of the starting graph - certain topological properties of the starting graph are conserved under duplication - while breaking the parent-daughter symmetry of the process by partial duplication induces non-conservation of this property and causes some "memory" of the starting graph to be lost. We find that full duplication does not produce power-law graphs, but partial duplication does. For partial duplication the power-law exponent depends, as the graph grows without bound, only on the growth process, and not on the starting graph.

A survey of existing results on scaling of large networks show a striking difference between biological and non-biological networks. Biological networks often have exponents that are between 1 and 2; that is, $1 < \beta < 2$ [8]. The non-biological networks , on the other hand, have exponents that commonly range from 2 to 4 or more (see Table I.). While it is difficult to draw a strong conclusion from this limited observation it does raise the question as to whether biological networks evolve differently. The non-biological networks have been convincingly modeled with preferential accretion of nodes [1-3 ], but these cannot explain exponents of power-law graphs less than 2.

**Table I.**

| | **NON-BIOLOGICAL** | |
|---|---|---|
| **Network** | **Approx. Exponent b** | **References** |
| Internet | 2.1 (in), 2.5 (out) | 1-6 |
| Citations | 3 | 6 |
| Actors | 2.3 | 6 |
| Power-grid | 4 | 1,6 |
| Phone calls | 2.1-2.3 | 6 |
| | **BIOLOGICAL** | |
| Yeast Protein-Protein Net | 1.5, 1.6, 1.7, 2.5 | 7, 9,24,25 |
| *E.coli* Metabolic Net | 1.7, 2.2 | 1, 10 |
| Yeast Gene Expression Net | 1.4-1.7 | 9 |
| Gene functional interactions | 1.6 | 11 |

Legend: Some examples of power-law distributions that have been examined and the exponent estimated for each of them. The references are indicated on the right. This is intended to be a representative sample of power-law behaviors.

A seminal idea in molecular evolution is that through gene duplication biological information is coopted, or “reused”, for different purposes [12]. This notion recognizes that the information in biomolecules, selected over hundreds of millions of years, represents a rich starting point for many useful modifications. This “reuse” occurs by the duplication and subsequent mutation of genes and other genetic elements, including both genes and *cis* regulatory sequences. The recent availability of genomic sequence information from a wide range of organisms provides abundant evidence of the widespread occurrence of gene duplication and the validity of the early hypotheses of Ohno and others. There is strong evidence, for example, that the genome of first eukaryote ever sequenced, that of a model organism, the yeast *Saccharomyces ceriviciae* (baker's yeast), is the result of an almost complete genome duplication in the distant past [16,17]. There is abundant evidence of duplication of large stretches of gene-containing DNA in humans [ 13,14], mice [15,16] and many other organisms [16-21]. In addition it is clear that many extensive "gene families" have evolved in which a basic amino acid sequence theme is used in modified form again and again for a variety of different purposes. Together with the variations that occur in parallel (including more complex processes like gene conversion), duplication provides the fundamental raw material that natural selection acts upon to evolve the genomes of living species, and is, therefore, a central process in the evolution genome-determined networks [22,23].

In our previous simulations of gene duplication [9] we found that duplication models in which all of the connections of the duplicated node are retained (full duplication models) do not exhibit a power-law distribution. On the other hand, modifications of the full duplication model in which only a fraction of the connections of each duplicated node are also duplicated and/or some of the duplicated connections are "re-wired", do exhibit power-law behavior. When the strong parent-daughter symmetry of the process of full duplication is broken by these modifications scale-free behavior emerges. In particular, partial duplication appears to reflect most of the observed properties of known biological networks [9].

**<u>Duplication Symmetry</u>**

We have focused on the invariant properties of graphs under the processes of duplication and have devised a representation that makes the invariant evident. By full duplication we mean the following. A random vertex, u, of graph $G_0$, that we call the sampling vertex, is selected. Then a new vertex v is added to $G_0$ in such a way that for each neighbor w of u, a new edge vw is added. This process is then repeated to evolve the graph under full duplication. The adjacency matrix , **A**, (n x n) determines the graph completely ( n is the number of vertices of the graph.) (see figure 1). However, we can use a more concise description which consists of a smaller matrix, **C**, derived from **A,** and a vector of integers. The reduced matrix, **C**, is obtained by removing all exactly repeated rows and columns to create a matrix made up only of unique rows and columns (if **A** is symmetric as for a non-directed graph, then so is **C**). The vector, **n**, is defined as the vector of the number of each kind of rows and columns removed to create **C**, plus 1– the number of identical rows and columns of each kind. It is evident that **A** can be reconstructed by adding back the number of rows and columns defined by **n -** thus the descriptions are equivalent. We define the *orbits* of the graph as subsets of nodes that are connected to exactly the same set of other nodes; that is, they have the same neighbors. Thus, orbits can be said to be equivalent to duplicated sets of nodes. **C** describes the connections between orbits - the "adjacency matrix for orbits". This description can be diagrammed as shown in an example in figure 2.

## Duplication

Now we consider a node duplication process in which each node has equal probability of being duplicated in a single time step (see Figure 1). The probability of randomly choosing a node for duplication in a particular orbit then, is equal to the fractional weighting of the orbit, as indicated by **n**. During growth by duplicating nodes at random, the probability of adding to a given orbit is simply the fraction of all nodes contained in that orbit, or $P(i) = \frac{\boldsymbol{n}_i}{\sum_j \boldsymbol{n}_j}$. If the node in orbit i is duplicated the vector element $\nu_i$ then increases by 1. As this process proceeds, the probability of duplication in any orbit is influenced by the history of prior duplications. This process results therefore in a "random asymptote" outcome in which the asymptotic occupation fraction of any orbit is a random variable. The degree of the nodes in any orbit, of course, does not change when a duplication occurs in that orbit, but the degrees of the some of the other orbits do change. This can be seen by explicitly writing the expression for the degree of the ith orbit in terms of the matrix C and the vector ν (where the sum is over all orbits).

$$k_i = \sum_j c_{i,j} \boldsymbol{n}_j \qquad \text{or} \qquad \vec{k} = C \cdot \boldsymbol{n} \tag{1}$$

In these terms the number of edges *s*, then, is simply

$$2s = \boldsymbol{n} \cdot \vec{k} = \boldsymbol{n} \cdot C \cdot \boldsymbol{n} \tag{2}$$

and $t = \sum_i \boldsymbol{n}_i$ , the total number of nodes of the graph.

The invariance, under duplication is now transparent. The dynamics of the duplication process in this description only changes the elements of **n**, the orbit populations, and leaves **C** entirely unchanged. Note that C is the adjacency matrix of the reduced sub-graph of the starting graph – it never changes under full duplication. The vector ***k***, like ***n***, changes under duplication, however, and is not an invariant.

Since the occupation numbers of the orbits, **n**, together with the invariant matrix **C** defines the graph completely, it is clear that the process of full duplication is commutative – it makes no difference in what order a defined set of nodes are duplicated. The history-independent "state vector" **n**, defines the graph with matrix **C**. Also it is clear that under duplication only the orbits present in the starting graph, $G_0$, are present in any of the progeny graphs, and we need only keep track of the changes of occupation of each orbit. This "memory" of the starting graph then is fully determined by **C.**

Note also that the chromatic number of a graph (the number of colors required to color all vertices while avoiding adjacent colors) is also conserved under duplication – it is invariant - and does not depend on the orbit occupation vector. Partial duplication also conserves the chromatic number. This can be seen by noting that removing edges after duplication cannot require re-coloring since the number of connected vertex pairs is reduced over full duplication, which is known to conserve the chromatic number.

In the full duplication process, the sizes of the orbits (i.e., the coordinates of **n)** , change at each time step. Assume that all orbits in the starting graphs have the same size, and suppose that $n$ is the number of orbits in the starting graph. If $t$ is the number of vertices at time $t$, the average orbit size is $a=t/n$. Then the distribution of the sizes of the orbits can be approximated by a density function $f(x)=e^{-x}$ for orbits of size $ax$ since the probability for an orbit having size $ax$ is proportional to $(1-x/n)^{n-1} \sim e^{-x}$ .

## The Partial Duplication Model

To consider partial duplication the network is again represented by a non-directed, un-weighted graph as above. Let $t_0$ be a constant and $G_{t_0}$ be a graph on $t_0$ vertices (see Figure 1). For $t>t_0$, $G_t$ is constructed by partial duplication from $G_{t-1}$ as follows. A random vertex, u, of $G_{t-1}$ , the sampling vertex, is selected. Then a new vertex v is added to $G_{t-1}$ in such a way that for each neighbor w of u, with probability p, a new edge v-w is added. The complete, or full duplication model results from setting $p=1$. Previous computational simulations indicate that that after many such partial duplications the degree distribution exhibits a power-law with an exponent, $\beta$ [9]. We show the following.

**Theorem 1.** *With probability approaching 1 as the number n of vertices becomes infinitely large, the partial duplication model with selection probability p generates power-law graphs with the exponent satisfying*

$$p(\beta - 1) = 1 - p^{\beta - 1} \qquad (3)$$

*In particular, if* $\frac{1}{2} < p < 1$ *then* $\beta < 2$.

The solutions for (3) that are illustrated in figure 3 consist of two curves. One is the line $\beta = 1$ (black). The other curve for $\beta$ is a montonically decreasing function of p (red). The two curve intersect at (x,1) where x = 0.56714329… the solution of $x=-\ln x$. One very interesting range for $\beta$ is when p is near ½. To get a power-law with exponent 1.5, for example, one should choose $p= 0.535898…$ This result is consistent with our previous simulation results for $p= ½$[8] . Also we see that the second curve for $\beta$ intersects zero at $p = \frac{\sqrt{5}-1}{2}$, which is an intriguing number (the "golden mean"). At $p$ = ½ one solution for $\boldsymbol{b}$ is 2. Although there are two solutions of $\boldsymbol{b}$ for each $p$, the stable solutions are on the red curve when $p <$ 0.56714329… and $\boldsymbol{b} = 1$ for $p >$ 0.56714329…. (A solution is considered *unstable* if the value of $f$ in the recurrence (5) below at a nearby point, $\boldsymbol{b} + \boldsymbol{e}$, diverges as indicated in the second ordered terms.) This is marked as the blue line in the figure.

Let us consider now a slightly more complex model in which a vertex may either be duplicated fully or partially.

**Theorem 2**. *With probability approaching 1 as the number of vertices becomes infinitely large, the mixed model, having full duplication with probability 1-q and partial duplication (with selection probability p) with probability q generates power-law graphs with the exponent* $\boldsymbol{b}$ *satisfying*

$$\boldsymbol{b}(1-q) + pq(\boldsymbol{b}-1) = 1 - qp^{b-1} \qquad (4)$$

To prove theorem 1 we need to establish a basic relationship between the number of vertices of specific degree at successive time steps. Recall that duplication starts adding one node per time step at $t_0$. Let $f(k,t)$ be the expected number of vertices with degree $k$ at time $t$. Its satisfies the following recurrence for $t \geq t_0$ :

$$f(k,t+1) = (1-\frac{p}{t})^k f(k,t) + (k-1)\frac{p}{t}(1-\frac{p}{t})^{k-1} f(k-1,t) + \sum_{j \geq k} \binom{j}{k} p^k (1-p)^{j-k} \frac{1}{t} f(j,t) \qquad (5)$$

The first term in the sum is due to those vertices of degree $k$ at time $t$ that are still vertices of degree $k$ at time $t+1$. The second term is due to those vertices of degree $k-1$ at time $t$, but of degree $k$ at time $t+1$. The third term is the expected value of a new vertex of degree $k$ that is generated at time $t+1$ expressed as a sum ranging over all $j$ where $j$ ³ $k$ (where a sampling vertex is chosen of degree $j$) A certain simplification can be achieved by letting $f(k,t) = a_k t + o(t)$. Namely, $a_k$ is essentially the fraction of vertices of degree $k$ at time $t$ - recall that one vertex is added each time step. Then $a_k$ must satisfy the following recurrence relation:

$$(1+kp)a_k = (k-1)pa_{k-1} + \sum_{j=k}^{\infty} \binom{j}{k} a_j p^k (1-p)^{j-k} \qquad (6)$$

Two lemmas will be useful in solving this recurrence relation, from which we can prove the theorem.

**Lemma 1**. *For a constant c and a real x that approaches infinity, we have*

$$\frac{\Gamma(x-c)}{\Gamma(x)} = (1+O(\frac{1}{x}))x^{-c} \quad and \quad \frac{\binom{x-c}{k}}{\binom{x}{k}} = (1+O(\frac{1}{x-k}))(1-\frac{k}{x})^{c}$$

*for all $k \le x$.*

**Lemma 2**. *For a fixed k, we have*

$$\sum_{j=k}^{\infty} \binom{j}{k} p^k (1-p)^{j-k} \left(\frac{k}{j}\right)^b = \left(1+O\left(\frac{1}{k}\right)\right) p^{b-1}$$

These two lemmas (whose proofs are found in the endnotes) can now be used to prove theorem 1.

**Proof of Theorem 1:**

We write $a_k = ck^{-b}$ for some $b$ and $c$ to be determined later. Using lemma 2 equation 6 can be rewritten as follows:

$$(1+kp-p^{b-1})k^{-b} = (k-1)p(k-1)^{-b} + O\left(k^{-b-1}\right) \qquad (7)$$

This implies that

$$1+kp-p^{b-1} = p\frac{(k-1)^{-b+1}}{k^{-b}}+O(1/k)$$

$$1-p^{b-1} = -p+bp+O(1/k)$$

Then by taking the limit as $k$ increases without bound, and defining $b$ at that limit as $\boldsymbol{b}$, we see that $\boldsymbol{b}$ satisfies the equation.

$$1-p^{b-1} = -p+\boldsymbol{b}p \qquad \text{or} \qquad p(\boldsymbol{b}-1)=1-p^{b-1}$$

which was to be proved.

**Proof of Theorem 2:**

Let $g(k,t)$ be the expected number of vertices with degree $k$ at time $t$ for the mixed duplication model. Its expected value satisfies the following recurrence for t ≥1 :

$$g(k,t+1)=(1-q)\left((1-\frac{1}{t})^k g(k,t)+(k-1)\frac{1}{t}(1-\frac{1}{t})^{k-1}g(k-1,t)+\frac{1}{t}g(k,t)\right)$$
$$+q\left((1-\frac{p}{t})^k g(k,t)+(k-1)\frac{p}{t}(1-\frac{p}{t})^{k-1}g(k-1,t)+\sum_{j\geq k}\binom{j}{k}p^k(1-p)^{j-k}\frac{1}{t}g(j,t)\right) \tag{8}$$

We set $g(k,t)=\boldsymbol{b}_k t+o(t)$ and $\boldsymbol{b}_k=ck^{-\boldsymbol{b}}$ and substitute into the above recurrence relation.

Then $\boldsymbol{b}$ satisfies

$$\boldsymbol{b}(1-q)+pq(\boldsymbol{b}-1)=1-qp^{b-1}.$$

## Discussion

We have investigated a node duplication model of network growth and present analytic results on the scaling of connectivities. This model, motivated by biological considerations, shows distinctively different scaling behavior from other models used to describe the growth of large networks, particularly the internet. For full duplication growth, we present a simple reduced matrix representation that captures the growth-invariant structure of the system. This way of looking at evolution under duplication makes it clear that the “memory” of the starting graph is retained under full duplication even though the nodes are chosen randomly for duplication. We demonstrate that network growth by full duplication does not result in power law distribution of connectivities, while partial duplication models do yield such behavior. Under partial duplication a significant part of the "memory" of the starting graph is lost. The distribution of degrees of the vertices, as shown by theorems 1 and 2, is independent of the starting graph, and the reduced matrix is no longer invariant. There is clearly some retained memory under partial duplication, however, in that the chromatic number of the graph, for example, is conserved, even under partial duplication.

Analytic relationships between the probability of edge duplication, $p$, in the partial duplication model, and the exponent of the resulting power law, ***b***, are presented in theorem 1. This simple relation shows that ***b*** is a monotonically decreasing function of $p$, for most of the domain $0 < p < 1$. Equation 3, however, has two solutions: ***b***=*1* and the more complex function shown in figure 3. For real graph evolution the intersection of the two curves represents a transition point. The generalization of these results, that includes the probability that a partial duplication occurs, $q$, is presented in theorem 2. The behavior of this model is also illustrated in Figure 3. The upper of the two curves for all values of $q$ appears to represent the actual solution, since these are the only stable solutions.

For the partial duplication model (Theorem 1) it is interesting that the range of significant interest of ***b***, between 1 and 2, is produced by only a relatively small range of selection probabilities p: $0.5 < p < 0.56714329\ldots$ Note that in this region, $2 > \beta > 1$, the dependence of ***b*** on $p$ is approximately linear: ***b*** ≈ 9.45 – 14.9 $p$. We find it curious that one of the solution curves intersects zero at the curious number *p = ( Ö5-1)/2*. This is the "golden mean", which is also the limit of the ratio of successive Fibonnaci numbers. In the mixed model, however, the solution curves also intersect for all $q < 1$, but the upper curve gives the stable solution with ***b*** ≥ 1 for all values of $p$.

What do our mathematical results have to do with biology? The sequencing of genomes in the past few years has made it clear that biological processes of evolution depend heavily on duplication of segments of the genome [13-21]. It is likely that the process of duplication is a major force in shaping real biological networks. While the abstraction of real biological networks studied here is unlikely to capture many properties of these biological networks, the global statistical properties of the networks and their topologies may be well represented by these kinds of models. It is certainly encouraging that the model analyzed here exhibits exponents of the degree distribution in the right range, and that the high cluster coefficients, like those seen for biological networks, are also seen[9]. Recent work in modeling segmentation in development [30] underlines the importance of the topology of regulatory networks. These authors made the intriguing observation that their model was rather insensitive to solution sets of the quantitative parameters of the model within large ranges of variation, but very sensitive to the topology of the network. This supports the idea that the statistical properties of the models considered here, which focus only on the connectivity of the network, are biological important.

While naturally occurring networks need not grow by a single mechanism, it may be possible to infer general growth and design principles from the global network properties of these networks. The comparisons made so far with non-biological networks, like the internet, show striking differences presumably because these examples evolve largely by non-biologically significant mechanisms. Successful models for internet-like structures [1-6] involve adding new nodes and connecting them preferentially to existing nodes of high degree. A striking difference between internet growth modes and biology is that biology only has previously defined relationships to work with (this information is carefully stored in the genome) while new web sites can be invented and attached to the net without copying previously invented web sites and their connections. If the internet evolved by random copying of previous web sites, we would expect a much more "biological" process, and one that could be modeled by duplication. In that case the exponent of the web power-law might be less than 2 rather than greater than 2 as observed. An important aspect of the differences between models is that the partial duplication model considered here can produce all values greater than 1 for scaling exponents, $\beta$, while the preferential connection models [1-4,8,28,29] can only produce exponents greater than 2. While many biological networks appear to exhibit power-law distributions with exponents between 1 and 2 (Table 1), this is not a constraint of the model, however, and some have been described with exponents greater than 2 [10]. The preferential connection models fail in biology for at least two reasons. They cannot explain the small exponents exhibited by some biological networks, but neither do they predict high clustering coefficients [9]. A ”copying” model for web growth has been proposed and analyzed by Kleinberg et al. [7]. While similar in some ways, this model is quite distinct from ours and predicts exponents greater than 2, like the preferential connection models.

The present simple model is a starting point from which more detailed and biologically accurate models can be derived. Inherent in all models that represent a biological network as a graph of this kind is that the strength of connections is ignored – the graph is unweighted. This leaves out some important complexity, for example whether a regulatory connection between nodes is positive or negative. Another limitation is the non-directed nature of the graphs considered here. We expect that, while somewhat more complex, the results for digraphs will reflect the same basic behaviors. The probability of duplication of edges is assumed to be uniform, which is another significant limitation in that in a real network some connections may well be much more important than others and therefore selection in real network will prefer to duplicate some edges more frequently than others. It is also likely that a certain amount of "re-wiring" takes place during growth; that is, connections of the new node to nodes that are not neighbors of the duplicated node can be made, and this is not considered in the present work. Finally, the absence of selection, a major driver in biological evolution, in these models is a limitation to be remedied in future work.

We conclude from our results and their limitations that, while more complex models will be needed, there is real biological content to the partial duplication model considered here that will likely provide some insights into the processes of evolution and perhaps something about the way in which biological networks function. A number of new problems suggested by these results, including extensions of the analysis mentioned above should provide a fruitful line of inquiry.

**<u>Acknowledgements</u>:** DG and GD wish to acknowledge a number of stimulating discussions with Ashish Bhan and Alpan Raval. This work was supported by NSF, NIH, the Norris foundation and the W.M. Keck Foundation.

## References :

1. Barabasi, A-L. and Albert, R. (1999) *Science* **286**, 509-512
2. Barabasi, A-L., Albert, R. and Jeong, H. (1999) *Physics A* **272**: 173-187
3. Albert, R. and Barabasi, A-L. (2002) *Rev. Mod. Phys*. **74**, 47-47
4. Albert, R., Jeong, H. and Barabasi,A-L. (1999) *Nature*, **401**, 130-131
5. Lu, L. *Proceedings of the 12th ACM-SIAM Symposium on Discrete Algorithms (SODA 2001),* 912-921 (2001)
6. Strogatz, S. *Nature*, **410**, 268-276 (2001)
7. Kleinberg, J., Kumar, S.R., Raghavan, P., Rajagopalan, S. and Tompkins, A. (1999) *Proc. Intl. Conf. On Combinatorics and Computing*, 1-18
8. Aiello, W., Chung, F. and Lu, L. (2000) *Proceeding of the 32nd Annual ACM Symposium on Theory of Computing*, 171-180
9. Bhan, A., Galas, D. and Dewey, D.G., *Bioinformatics* (in press)
10. Jeong, H., Tombor, B., Albert, R., Oltvai, Z.N. and Barabasi, A-L. (2000) *Nature* **407**, 651-654
11. Snel, B., Bork, P. and Huynen, M. A., (2002) *Proc. Nat. Acad. USA*, **99**, 5890-5895
12. Ohno, S. (1970) *Evolution by Gene Duplication*, Springer Verlag, New York
13. Venter, J. C. et al. (2001) *Science* **291,**1304-1351
14. International Human Genome Sequencing Consortium. (2001) *Nature* **409**, 860-921
15. Dehal, P., Predki, P., Olsen, A. S., Kobayashi, A., Folta, P., Lucas, S., Land, M. , Terry, A., *et al*., (2001) *Science* **293**, 104-111

16. Stubbs, L. (2002) in *Genomic Technologies: Present and Future*, (Galas, D. and McCormack,S. eds.) pp.43-66, Caister Academic Press
17. Friedman, R. and Hughes, A. (2002) *Genome Research* 11: 373-381
18. Wolfe, K. H. and Shields, D.C. (1997) *Nature* **387**, 708-713
19. Seioghe, C. and Wolfe, K.H. (1999) *Gene* **238**, 253-261
20. Sidow, A. (1996) *Curr. Opinion Genet. Dev*. **8**, 694-700
21. Gu, Z. Cavalcanti, A., Chen, F-C, Bouman, P. and Li, W-H. (2002) *Mol. Biol and Evol.* **19**: 256-262
22. Wagner, A. (1994) *Proc. Nat Acad. Sci.* USA, **91**, 4387-4391
23. Wagner, A. (2001) *Mol. Biol. Evol.* **18**, 1283-1292.
24. Dewey, T.G., Galas, D. (2001) *Func. Integr. Genomics* **1**, 269-278.
25. Dorogovtsev, S.N., and Mendes, J.F.F. (2001) *Phys. Rev. E.* **63**, 056125, 1-18
26. Ito, T., Chiba, T., Ozawa, R., Toshida, M., Hattori, M. and Sakaki, Y. (2001) *Proc. Nat. Acad. Sci. USA* **98**, 4569-4574
27. Uetz, P., Giot, L. Cagney, G. Mansfield, T., Judson, R. Knight, J. Lockshon, D. et. Al. (2000) *Nature*, **403**, 623-627
28. Krapivsky, P.L., Redner, S. and Leyvraz, F. (2000) *Phys Rev. Lett.* , **85**, 46-29-4632
29. Rzhetsky, A. and Gomez, S.M. (2001). *Bioinformatics*, **17**, 988-996
30. Von Dassow, G. Meir, E. Munro, E.M. and Odell, G.M. (2000) *Nature* **406**, 188-192

***<u>End note 1</u>***

**Proof of lemma 1:**

Since we have

$$\Gamma(x) = \sqrt{\frac{2p}{x}}\left(\frac{x}{e}\right)^x\left(1+\frac{1}{12x}+O\left(\frac{1}{x^2}\right)\right)$$

then

$$\frac{\Gamma(x-c)}{\Gamma(x)} = (1+O\left(\frac{1}{x}\right))\sqrt{\frac{2p\,/(x-c)}{2p\,/\,x}}\,,\frac{\left(\frac{x-c}{e}\right)^{x-c}}{\left(\frac{x}{e}\right)^{x}}$$

$$= \left(1+O\left(\frac{1}{x}\right)\right)x^{-c}$$

Also

$$\frac{\binom{x-c}{k}}{\binom{x}{k}} = \frac{\Gamma(x-c+1)\,/\,\Gamma(x-c+1-k)}{\Gamma(x+1)\,/\,\Gamma(x+1-k)}$$

$$= \left(1+O\left(\frac{1}{x-k}\right)\right)\frac{x^{-c}}{(x+1-k)^{-c}} = \left(1+O\left(\frac{1}{x-k}\right)\right)\left(1-\frac{k}{x}\right)^{c}$$

which proves lemma 1.

***End note 2***

**Proof of lemma 2:** Using lemma 1 we write

$$\sum_{j=k}^{\infty}\binom{j}{k}p^{k}(1-p)^{j-k}\left(\frac{k}{j}\right)^{b} = \sum_{j=k}^{\infty}\binom{j}{j-k}p^{k}(1-p)^{j-k}\left(\frac{k}{j}\right)^{b}$$

$$= \left(1+O\left(\frac{1}{k}\right)\right)\sum_{j=k}^{\infty}\binom{j-b}{j-k}p^{k}(1-p)^{j-k}$$

$$= \left(1+O\left(\frac{1}{k}\right)\right)p^{k}\sum_{m=0}^{\infty}\binom{m+k-b}{m}(1-p)^{m}$$

$$= \left(1+O\left(\frac{1}{k}\right)\right)p^{k}\sum_{m=0}^{\infty}\binom{b-k-1}{m}(-1)^{m}(1-p)^{m}$$

$$= \left(1+O\left(\frac{1}{k}\right)\right)p^{k}p^{b-k-1} = \left(1+O\left(\frac{1}{k}\right)\right)p^{b-1}$$

and lemma 2 is proved.

**Figures Legends:**

**Figure 1:** A diagram illustrating the process of node duplication and partial duplication as described in the text. The nodes are labeled as referred to in the text.

**Figure 2**: A graph, shown on the left, can be described as a set of orbits (middle diagram) connected as shown on the right**. C** is the adjacency matrix of the right graph, **n** is the vector of the occupation numbers. The right most panel shows how the vertices can be grouped into the 4 orbits of the graph, thus the $\nu$ vector is simply (4, 4, 1, 2).

**Figure 3**: Solutions for the equations of theorems 1 and 2. The graph of the exponent $\beta$ as a function of p from theorem 1 is plotted as the red line in this plot (same as theorem 2 with q=1). The zero for the solution curve is indicated. Curves for two other values of q (the probability of partial duplication of a vertex.) in the equation of Theorem 2 are in green, q=0.5, and blue, q=0.8. $\beta$=1 (in black) is always a solution for any q. The stable solution is always above $\beta$=1 (including $\beta$=1 for p large enough) The aqua-colored line shows the stable solutions for q=1.

Figure 1

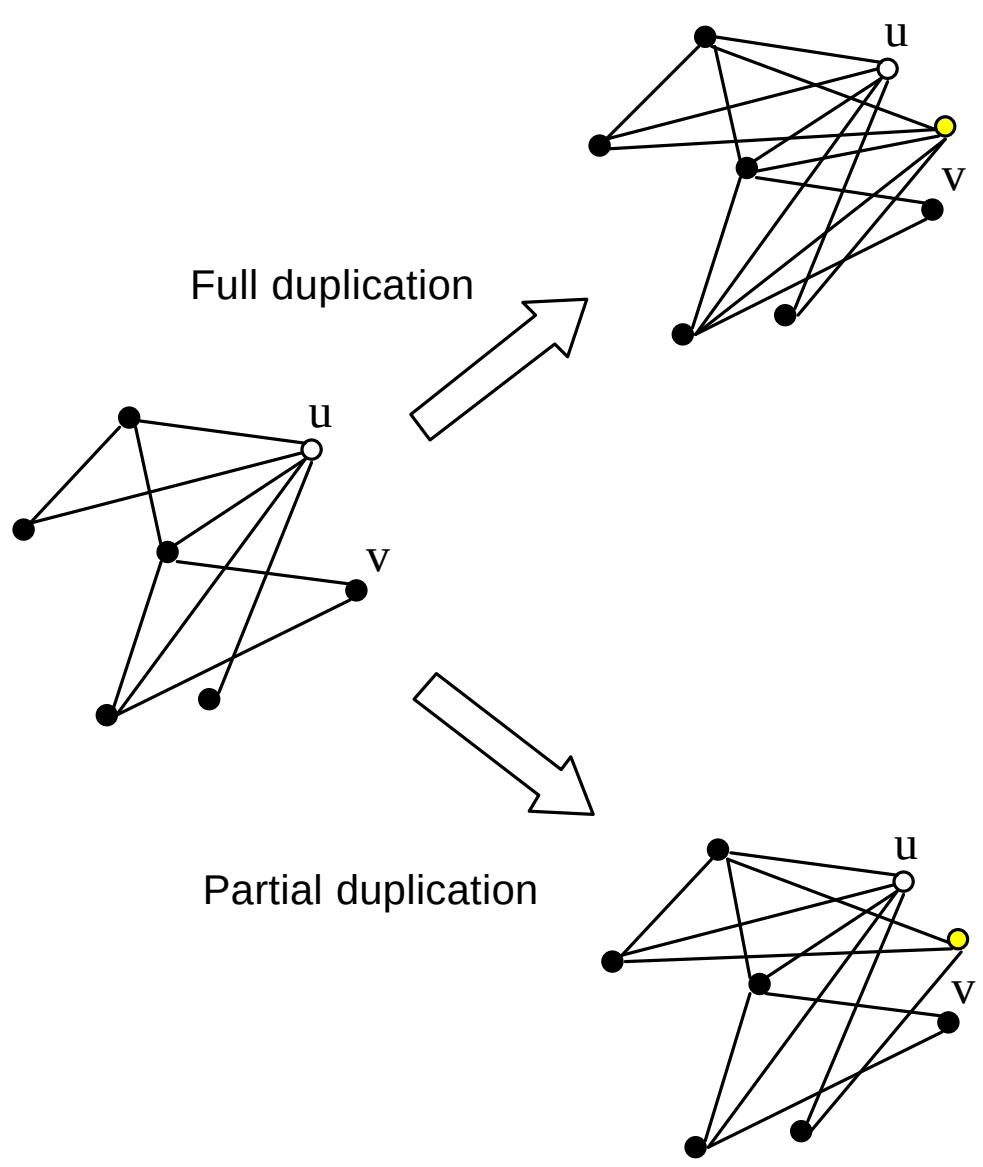

Figure 2

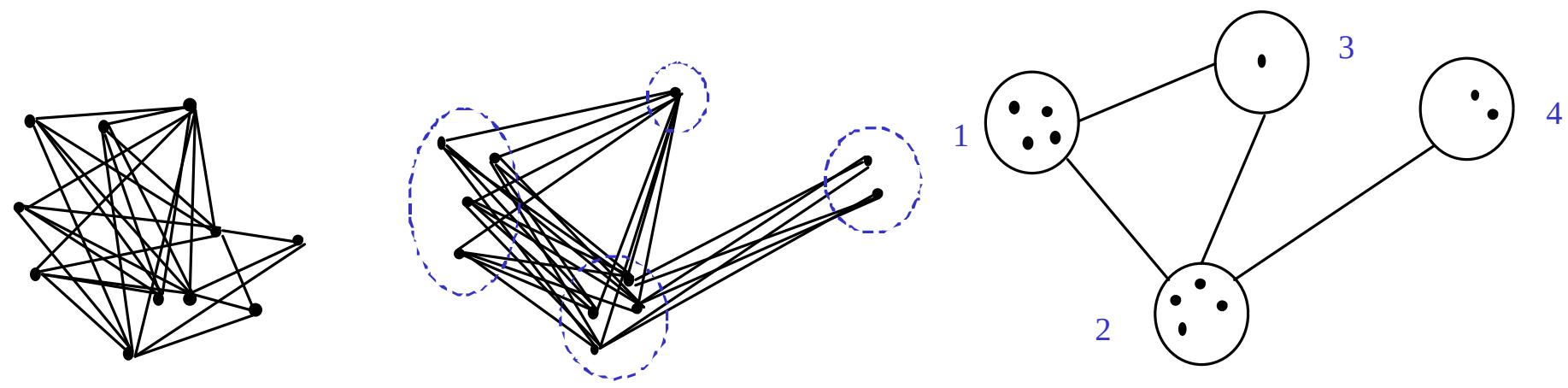

Figure 3

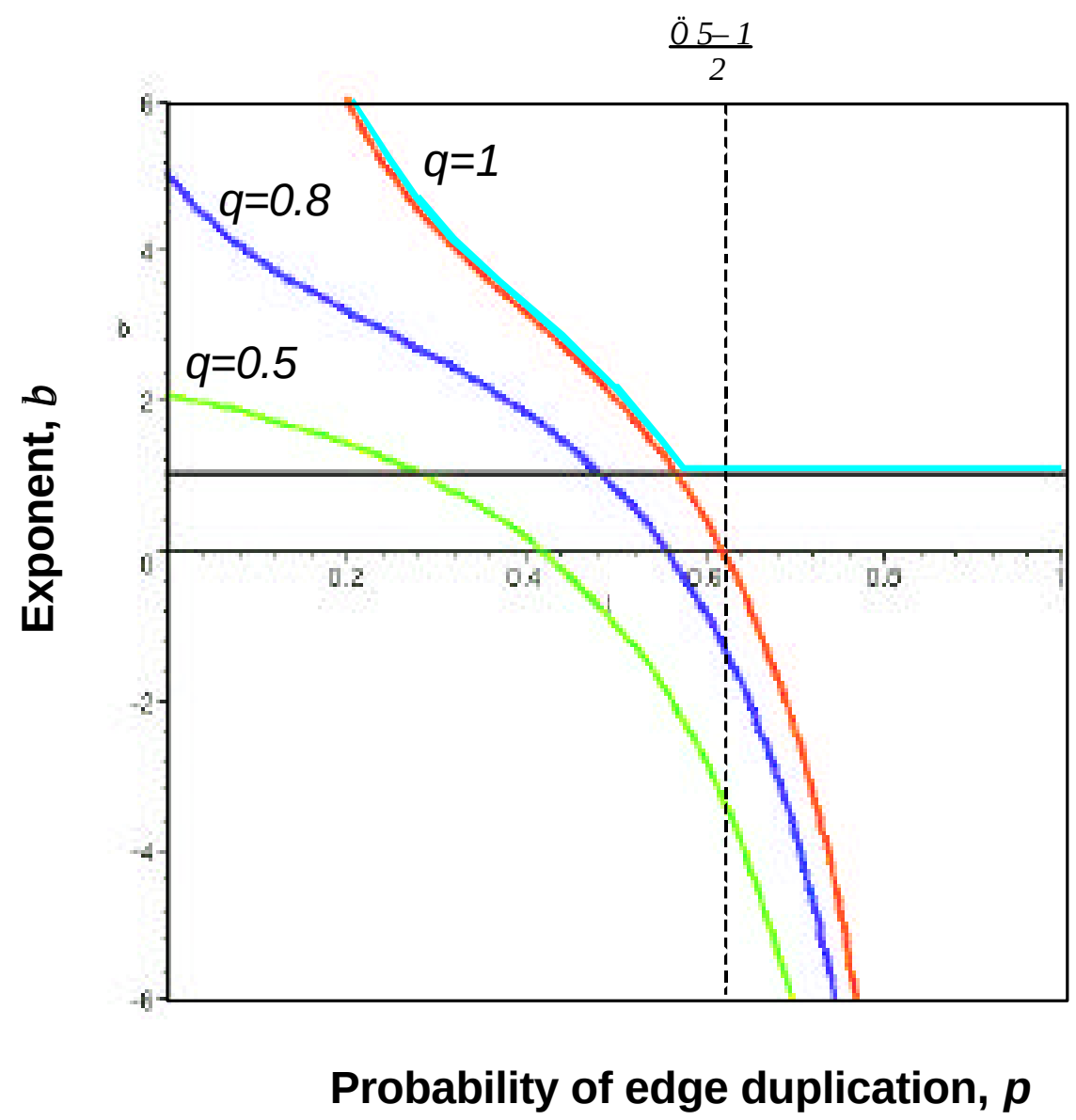
Ö 5– 1
2
q=1
q=0.8
q=0.5
Exponent, b
0.2
0.4
0.6
0.8
Probability of edge duplication, p